\documentclass[12pt,english,floatfix,superscriptaddress,aps,preprint,amsmath,amssymb]{revtex4}
\usepackage[latin9]{inputenc}
\usepackage{amsmath}
\usepackage{amssymb} 
\usepackage{amsbsy}
\usepackage{amsfonts}
\usepackage{amsopn}
\usepackage{amstext}
\usepackage{graphicx}
\usepackage{esint}
\usepackage{hyperref}

\usepackage{amssymb}
\usepackage{amsfonts}
\usepackage{amsmath}
\usepackage{graphicx}
\usepackage[english]{babel}
\usepackage{color}
\usepackage[dvips]{epsfig}
\usepackage[dvips]{graphicx}
\usepackage{float}
\usepackage{units}
\usepackage{textcomp}
\usepackage{babel}

\def \be {\begin{equation}}
\def \ee {\end{equation}}
\def \bea{\begin{eqnarray}}
\def \eea{\end{eqnarray}}
\def \ba {\begin{align}}
\def \ea {\end{align}}

\def \a {\alpha}
\def \b {\beta}
\def \g {\gamma}

\def \d {\delta}

\def \e {\epsilon}

\def \m {\mu}
\def \n {\nu}

\def \l {\lambda}

\def \s {\sigma}

\def \r {\rho}

\def \th {\theta}

\def \t {\tau}

\def \st {\star}

\def \f {\frac}

\def \nn {\nonumber}

\def \la {\leftarrow}

\def \la {\label}

\begin{document}

\title{Generalized Hodge dual for torsion in teleparallel gravity}

\author{Peng Huang}
\email{huangp46@mail.sysu.edu.cn}
\affiliation{School of Astronomy and Space Science, Sun Yat-Sen University, Guangzhou
510275, China}

\author{Fang-Fang Yuan}
\email{ffyuan@nankai.edu.cn}
\affiliation{School of Physics, Nankai University, Tianjin 300071, China}

\begin{abstract}

For teleparallel gravity in four dimensions,
Lucas and Pereira have shown that a generalized Hodge dual for torsion tensor can be defined with coefficients determined by mathematical consistency.
In this paper, we demonstrate that
a direct generalization to other dimensions fails and no new generalized Hodge dual operator could be given.
Furthermore,
if one enforces the definition of a generalized Hodge dual to
be consistent with the action of teleparallel gravity in general dimensions,
the basic identity for any sensible Hodge dual
would require an \textit{ad hoc} definition for the second Hodge dual operation which is totally unexpected.
Therefore, we conclude that at least for the torsion tensor, the observation of
Lucas and Pereira only applies to four dimensions.

\end{abstract}
\maketitle

\section{Introduction}

Teleparallel gravity is a gauge theory for the translation group. Although the central role in this theory is played by torsion rather than curvature, it is in fact equivalent to general relativity \cite{Aldrovandi:2013wha}. In contrast with those internal (Yang-Mills-type) gauge theories,
teleparallel gravity is generally associated with soldered bundles
where the spacetime (external) and gauge (internal) indices can be transformed into
each other through tetrad fields $e^a_{\m}$. Due to this property, one can construct more types of scalar from torsion tensor than the usual internal gauge theories where no transformation between external and internal indices happens. For example, in internal gauge theories, a scalar can be constructed as $F^{A\m\n}F_{A\m\n}$ with $A$ the internal index and $\m\n$ the spacetime indices; however, in teleparallel gravity, in addition to term $T^{a\m\n}T_{a\m\n}=T^{a\m\n}e_{a}^{\l}T_{b\m\n}e^b_{\l}=T^{\l\m\n}T_{\l\m\n}$ that corresponds to $F^{A\m\n}F_{A\m\n}$, terms like $T^{a\m\n}T_{\m a\n}=T^{\l\m\n}T_{\m\l\n}$ and $T^{\m a}_{\ \ \ \m}T^{\n}_{\ \ a\n}=T^{\m\l}_{\ \ \ \m}T^{\n}_{\ \ \l\n}$ are also possible.

The appearance of new scalar terms in teleparallel gravity makes the definition of Hodge dual for torsion tensor, the gauge field strength of the translation group, a nontrivial question. Nevertheless, in the case of $D=3+1$ teleparallel gravity,
Lucas and Pereira \cite{Lucas:2008gs} (partially based on the investigation in \cite{deAndrade:2005xy})
have proposed an approach to find this generalization for the torsion tensor.  In their approach, similar to the case of Yang-Mills theories in $D=3+1$ spacetime where the Hodge dual of Yang-Mills field strength $F^A_{\ \m\n}$ is defined as $(\star F )^A_{\ \m\n}=\f{1}{2} h \e_{\m\n\r\s}F^{A\r\s}$ \footnotemark [1] \footnotetext[1]{In this definition, we write the parentheses explicitly to remind the reader that the indices $^A_{\ \m\n}$ denote the indices of the Hodge dual of field strength, $\star F$, which should not be confused with the indices $^A_{\ \m\n}$ in the field strength $F$ in $F^A_{\ \m\n}$. }, they speculated that the generalized Hodge dual for torsion in $D=3+1$ should have the form as
\be \label{h4.1}
(\st T )^{\l}_{\ \m\n} = h \e_{\m\n\r\s} \Big(\f{a}{2}  T^{\l\r\s} + a T^{\r\l\s}  + c T^{\th\r}_{\ \ \th} g^{\l\s} \Big),
\ee
the first term is just as that in Yang-Mills theories, the second and third term appear due to new possibilities of index contraction, furthermore, the factor $\f{1}{2}$ in the first term is needed to remove equivalent terms of the summation. Then, for the generalized Hodge dual of a $k$-form to be consistent, it should obey the following property
\be
\label{23}
\st \st A^{k} = (-1)^{k(D-k)+(D-s)/2} A^{k},
\ee
with $D$ the dimension of the spacetime and $s$ the metric signature, in the case of $D=4$ and $s=2$, (\ref{23}) implies that
\be
\label{24}
(\st \st T)^{\l}_{\ \m\n} = - T^{\l}_{\ \m\n}.
\ee
Roughly speaking, (\ref{h4.1}) and (\ref{24}) tell that, with appropriate factors, the operation of Hodge dual square maps differential form back to itself. This key constraint will eventually enable one to determine the unknown coefficients in definition (\ref{h4.1}). In this way, a consistent definition for the generalized dual torsion in $D=3+1$ has been found in \cite{Lucas:2008gs}.

The primary motivation of the present work is to extend the above idea to general dimensions with $D\geq 2+1$. We will show that, the extension is not so straightforward as expected, in fact, no solution will exist for the constraint equation, see  (\ref {50}) and  (\ref {66}) below. We ultimately find that two basic requirements suffice to lead to a new definition which is applicable to general dimensional teleparallel gravity.
Firstly, one still has to impose the key property  (\ref{23}) of a consistent Hodge dual. 
Secondly, we demand that starting with the wedge product of the torsion tensor and its generalized Hodge dual, the standard Lagrangian of teleparallel gravity should be naturally reproduced.
The caveat is that
the new definition involves a specific second Hodge dual operation
whose mathematical reliability and physical importance need to be critically inspected.

The structure of this paper is as follows.
In the next section, we explain the procedure of \cite{Lucas:2008gs} in detail to obtain the $D=3+1$ generalized Hodge dual torsion.
We then explicitly show in section \ref{3d} that a straightforward generalization fails in $D=2+1$.
In section \ref{nd}, a possible solution to this problem is given for the teleparallel gravity in general dimensions.
We conclude our discussion in section \ref{con}.

\section{Generalized Hodge dual for torsion in $D=3+1$: A detailed derivation}     \la{4d}

Since we aim to extend the method of \cite{Lucas:2008gs} (see also Chapter 8 of \cite{Aldrovandi:2013wha}) to general dimensions,
it would be instructive to fill in the detail of this process as outlined there.
This will prove to be necessary especially when one realizes
that the generalization turns out to be not so straightforward. To streamline the exposition, we denote the generalized Hodge dual $(\st T)^{\l}_{\ \m\n}$ as $B^{\l}_{\ \m\n}$.
Thus according to the definition in  (\ref{h4.1}), we have
\be
\label{26}
B^{\l\g\d} =  \f{1}{h}\e^{\g\d\a\b} \Big( \f{a}{2}T^{\l}_{\ \a\b} + a T^{\ \l}_{\a\ \b} + c T^{\t}_{\ \a\t} g^{\l}_{\b} \Big),
\ee
\be
\label{27}
B^{\g\l\d} = \f{1}{h}\e^{\l\d\a\b} \Big( \f{a}{2}T^{\g}_{\ \a\b} + a T^{\ \g}_{\a\ \b} + c T^{\t}_{\ \a\t} g^{\g}_{\b} \Big),
\ee
\be
\label{28}
B^{\th\g}_{\ \ \th} =  \f{1}{h}\e^{\g\th\a\b} \Big( \f{a}{2}T_{\th\a\b} + a T_{\a\th\b} + c T^{\t}_{\ \a\t} g_{\th\b} \Big).
\ee
On the other hand, we have
\be
\label{29}
(\st \st T)^{\l}_{\ \pi\chi}=(\st B)^{\l}_{\ \pi\chi}=h \e_{\pi\chi\g\d} \Big(\f{a}{2} B^{\l\g\d} + a B^{\g\l\d}  + c B^{\th\g}_{\ \ \th} g^{\l\d} \Big).
\ee
Inserting  (\ref{26})-(\ref{28}) into  (\ref{29}), we arrive at
\bea
\label{30}
(\st \st T)^{\l}_{\ \pi\chi}=(\st B)^{\l}_{\ \pi\chi} &=& h \e_{\pi\chi\g\d}\bigg [ \f{a}{2h}\e^{\g\d\a\b} \Big( \f{a}{2} T^{\l}_{\ \a\b} + a T^{\ \l}_{\a\ \b} + c T^{\t}_{\ \a\t} g^{\l}_{\b} \Big)  \nn   \\
&&+  \f{a}{h}\e^{\l\d\a\b} \Big(\f{a}{2} T^{\g}_{\ \a\b} + a T^{\ \g}_{\a\ \b} + c T^{\t}_{\ \a\t} g^{\g}_{\b} \Big)     \nn  \\
&& +  \f{c}{h}\e^{\g\th\a\b} \Big( \f{a}{2} T_{\th\a\b} + a T_{\a\th\b} + c T^{\t}_{\ \a\t} g_{\th\b} \Big)g^{\l\d}\bigg ] \nn  \\
&=&  \e_{\pi\chi\g\d}\e^{\g\d\a\b} \Big(\f{a^2}{4} T^{\l}_{\ \a\b} + \f{a^2}{2} T^{\ \l}_{\a\ \b} + \f{ac}{2} T^{\t}_{\ \a\t} g^{\l}_{\b} \Big)  \nn   \\
&&+ \e_{\pi\chi\g\d}\e^{\l\d\a\b} \Big(\f{a^2}{2} T^{\g}_{\ \a\b} + a^2 T^{\ \g}_{\a\ \b} + ac T^{\t}_{\ \a\t} g^{\g}_{\b} \Big)     \nn  \\
&& + \e_{\pi\chi\g\d}\e^{\g\th\a\b} \Big( \f{ac}{2} T_{\th\a\b} + ac T_{\a\th\b} + c^2 T^{\t}_{\ \a\t} g_{\th\b} \Big)g^{\l\d}.
\eea
The general formula for the products of Levi-Civita symbols reads
\be
\label{999}
\e_{i_1 ... i_k \ i_{k+1} ... i_n} \e^{i_1 ... i_k \ j_{k+1} ... j_D} = (-1)^{(D-s)/2} k ! (D-k) !\ \d^{[j_{k+1}}_{i_{k+1}} ... \d^{j_D]}_{i_D}
\ee
with $D$ the spacetime dimension and $s$ the signature of the metric (in present case $D=4$ and $s=2$).
Taking advantage of this fact, we get the following useful expressions
\bea
\label{eee}
\e_{\pi\chi\g\d}\e^{\g\d\a\b} &=& -2 \big( \d^{\a}_{\pi} \d^{\b}_{\chi} - \d^{\b}_{\pi} \d^{\a}_{\chi} \big),   \\
\label{rrr}
\e_{\pi\chi\g\d}\e^{\l\d\a\b} &=& -  \big( \d^{\l}_{\pi} \d^{\a}_{\chi} \d^{\b}_{\g} + \d^{\a}_{\pi} \d^{\b}_{\chi} \d^{\l}_{\g} + \d^{\b}_{\pi} \d^{\l}_{\chi} \d^{\a}_{\g}  - \d^{\a}_{\pi} \d^{\l}_{\chi} \d^{\b}_{\g} - \d^{\b}_{\pi} \d^{\a}_{\chi} \d^{\l}_{\g} - \d^{\l}_{\pi} \d^{\b}_{\chi} \d^{\a}_{\g}  \big),    \\
\label{ttt}
\e_{\pi\chi\g\d}\e^{\g\th\a\b}  &=& - \big( \d^{\th}_{\pi} \d^{\a}_{\chi} \d^{\b}_{\d} + \d^{\a}_{\pi} \d^{\b}_{\chi} \d^{\th}_{\d} + \d^{\b}_{\pi} \d^{\th}_{\chi} \d^{\a}_{\d} - \d^{\a}_{\pi} \d^{\th}_{\chi} \d^{\b}_{\d} - \d^{\b}_{\pi} \d^{\a}_{\chi} \d^{\th}_{\d} - \d^{\th}_{\pi} \d^{\b}_{\chi} \d^{\a}_{\d}  \big),
\eea
it's apparent that the spacetime dimension $D$ will come into the final result from term $\e_{\pi\chi\g\d}\e^{\l\d\a\b}\cdot ac T^{\t}_{\ \a\t} g^{\g}_{\b}$.

Then, with  (\ref{eee})-(\ref{ttt}), we can simplify (\ref{30}) to
\bea
\label{31}
(\st \st T)^{\l}_{\ \pi\chi} &=& - (2a^2-ac) T^{\l}_{\ \pi\chi} - (2a^2+ac) \big(  T^{\ \l}_{\pi\ \chi} - T^{\ \l}_{\chi\ \pi} \big)  \nn  \\
&&- \big[ 2a^2+(D-3)ac \big] \big( T^{\t}_{\ \chi\t} \d^{\l}_{\pi} - T^{\t}_{\ \pi\t} \d^{\l}_{\chi} \big).
\eea

By recalling the basic property of a Hodge dual operator as in (\ref{24}), we get the following constraint equations
\be
\label{32}
\begin{cases}
2a^2-ac = 1, \\
2a^2+ac = 0, \\
2a^2+(D-3)ac = 0.
\end{cases}
\ee
In $D=3+1$ case, the problem boils down to solving two equations as
\be
\label{33}
\begin{cases}
2a^2-ac = 1, \\
2a^2+ac = 0. \\
\end{cases}
\ee
The solution can be found to be $a=\f{1}{2}$, $c=-1$.
Note that a generalization of the usual expression $(\star F )^A_{\ \m\n}=\f{1}{2} h \e_{\m\n\r\s}F^{A\r\s}$
requires $a$ to be positive.
So we have discarded the other solution $a=-\f{1}{2}$, $c=1$ as in \cite{Lucas:2008gs}.

All in all, the generalized Hodge dual torsion we are looking for is
\be   \la{h4d}
(\st \ T)^{\l}_{\ \m\n} = h \e_{\m\n\r\s} \Big( \f{1}{4} T^{\l\r\s} +\f{1}{2} T^{\r\l\s}-T^{\th\r}_{\ \ \th} g^{\l\s} \Big),
\ee
and a consistency check is that this definition can be used to reproduce the Lagrangian of teleparallel gravity.

It should be stressed that although one necessarily has $D=3+1$ here, we retain the constant $D$ in the above derivation to track that
how the spacetime dimension comes into play.
By closely examining the $D=2+1$ case in the subsequent section, it will become clear that
the main obstacle to a straightforward generalization of the above procedure
could be precisely attributed to the nontrivial role 
of the spacetime dimension.

\section{The $D=2+1$ case: A naive approach and its problem}  \la{3d}

Our work grows out of an effort
to study the properties of a generalized Hodge dual in the $D=2+1$ case
which we expatiate upon in this section. By analogy with the previous discussion,
a generalized Hodge dual for torsion in $D=2+1$
could be defined as
\be
\label{41}
(\st T)^{\l}_{\ \m} = h \e_{\m\r\s} \Big(\f{a}{2} T^{\l\r\s} + a T^{\r\l\s}  + c T^{\th\r}_{\ \ \th} g^{\l\s} \Big).
\ee
As before, the relative factor in front of the first two terms is necessary to remove
equivalent terms coming from the summation.
By denoting $(\st T)^{\l}_{\ \m}$ as $B^{\l}_{\ \m}$,
we have
\bea
\label{43}
B^{\l\g} &=&  \f{1}{h}\e^{\g\a\b} \Big( \f{a}{2} T^{\l}_{\ \a\b} + a T^{\ \l}_{\a\ \b} + c T^{\t}_{\ \a\t} g^{\l}_{\b} \Big),   \\
\label{44}
B^{\g\l} &=& \f{1}{h}\e^{\l\a\b} \Big( \f{a}{2}T^{\g}_{\ \a\b} + a T^{\ \g}_{\a\ \b} + c T^{\t}_{\ \a\t} g^{\g}_{\b} \Big),   \\
\label{45}
B^{\th}_{\ \th} &=&  \f{1}{h}\e^{\th\a\b} \Big( \f{a}{2} T_{\th\a\b} + a T_{\a\th\b} + c T^{\t}_{\ \a\t} g_{\th\b} \Big).
\eea
On the other hand, we have
\be
\label{46}
(\st \st T)^{\l}_{\ \pi\chi}=(\st B)^{\l}_{\ \pi\chi}=h \e_{\pi\chi\g} \big( a B^{\l\g} + a B^{\g\l}  + c B^{\th}_{\ \th} g^{\l\g} \big).
\ee
Because of the absence of equivalent terms of the summation here,
one has the same factors in front of the first two terms which is totally different from the case in $D=3+1$.

Inserting (\ref{43}), (\ref{44}) and (\ref{45}) into (\ref{46}), one can find that two successive operations of generalized Hodge dual lead to
\bea
\label{47}
(\st \st T)^{\l}_{\ \pi\chi}=(\st B)^{\l}_{\ \pi\chi} &=& h \e_{\pi\chi\g}\bigg [ \f{a}{h}\e^{\g\a\b} \Big( \f{a}{2} T^{\l}_{\ \a\b} + a T^{\ \l}_{\a\ \b} + c T^{\t}_{\ \a\t} g^{\l}_{\b} \Big)  \nn   \\
&&+ \f{a}{h}\e^{\l\a\b} \Big(\f{a}{2} T^{\g}_{\ \a\b} + a T^{\ \g}_{\a\ \b} + c T^{\t}_{\ \a\t} g^{\g}_{\b} \Big)     \nn  \\
&& +  \f{c}{h}\e^{\th\a\b} \Big( \f{a}{2}  T_{\th\a\b} + a T_{\a\th\b} + c T^{\t}_{\ \a\t} g_{\th\b} \Big) g^{\l\g}\bigg ] \nn  \\
&=& - (2a^2-ac) T^{\l}_{\ \pi\chi} - (2a^2+ac) \big(  T^{\ \l}_{\pi\ \chi} - T^{\ \l}_{\chi\ \pi} \big)  \nn  \\
&&- \big[ 2a^2+(D-3)ac \big] \big( T^{\t}_{\ \chi\t} \d^{\l}_{\pi} - T^{\t}_{\ \pi\t} \d^{\l}_{\chi} \big).
\eea

For the $D=2+1$ case here, we have $D=3, s=1$, thus  (\ref{23}) implies that $(\st \st T)^{\l}_{\ \m\n} = - T^{\l}_{\ \m\n}$.
Therefore, 
the constraint equations for the coefficients are given as
\be
\label{49}
\begin{cases}
2a^2-ac = 1, \\
2a^2+ac = 0, \\
2a^2+(D-3)ac = 0.
\end{cases}
\ee
This means that we have to solve the equation set 
\be
\label{50}
\begin{cases}
2a^2-ac = 1, \\
2a^2+ac = 0, \\
2a^2=0.
\end{cases}
\ee
Since no solution could be found, we arrive at the conclusion that
a naive implementation of the approach in \cite{Lucas:2008gs} does not work for $D=2+1$ case.
In fact, the constraint equations will always have no sensible solutions except for the four dimensional case.  
However, as we shall show in the next section, a new generalized Hodge dual could indeed be given for the torsion tensor in general dimensions. The new definition will also be in accord with the standard Lagrangian of teleparallel gravity.

\section{General dimensions: Towards a new definition}
                                                    \la{nd}
\subsection{A direct generalization}

For the torsion tensor in general dimensions with Lorentzian signature, we start with the following definition of a generalized Hodge dual
\be
\label{61}
(\st T)^{\l}_{\ \a_1...\a_{D-2}}=h\e_{\a_1...\a_{D-2}\m\n} \Big( \f{a}{2}T^{\l\m\n} +aT^{\m\l\n}+cT^{\th\m}_{\ \ \th} g^{\l\n} \Big).
\ee
It can be easily checked that this definition go back to  (\ref{h4.1}) when the spacetime comes to the $D=3+1$ case.
Accordingly we have
\bea
\label{62}
(\st T)^{\l\a_1...\a_{D-2}} &=& \f{1}{h}\e^{\a_1...\a_{D-2}\m\n} \Big (\f{a}{2}T^{\l}_{\ \m\n} +aT^{\ \l}_{\m\ \n}+cT^{\th}_{\ \m\th} g^{\l}_{\n}\Big ),  \\
\label{63}
(\st T)^{\a_1\l...\a_{D-2}} &=& \f{1}{h}\e^{\l...\a_{D-2}\m\n} \Big (\f{a}{2}T^{\a_1}_{\ \ \m\n} +aT^{\ \a_1}_{\m\ \ \n}+cT^{\th}_{\ \m\th} g^{\a_1}_{\n}\Big ),  \\
\label{64}
(\st T)_{\th}^{\ \a_1\th\a_3...\a_{D-2}} &=& \f{1}{h}\e^{\a_1\th\a_3...\a_{D-2}\m\n} \Big (\f{a}{2} T_{\th\m\n} +aT_{\m\th\n}+cT^{\r}_{\ \m\r} g_{\th\n}\Big ).
\eea
Two successive operations of generalized Hodge dual lead to
\bea
\label{65}
(\st \st T )^{\l}_{\ \a\b}&=&[\st (\st T)]^{\l}_{\ \a\b}  \nn \\
&=&h\e_{\a\b\a_1...\a_{D-2}} \bigg[ (\st T)^{\l\a_1...\a_{D-2}}\cdot a \cdot \f{1}{(D-2)!}+(\st T)^{\a_1\l...\a_{D-2}}\cdot a\cdot\f{1}{(D-3)!} \nn \\
&&+(\st T)_{\th}^{\ \a_1\th\a_3...\a_{D-2}}g^{\l\a_2}\cdot c\cdot\f{1}{(D-3)!} \bigg] \nn \\
&=&h\e_{\a\b\a_1...\a_{D-2}}\bigg[ \f{1}{h}\e^{\a_1...\a_{D-2}\m\n} \Big (\f{a}{2}T^{\l}_{\ \m\n} +aT^{\ \l}_{\m\ \n}+cT^{\th}_{\ \m\th} g^{\l}_{\n}\Big )\cdot a \cdot \f{1}{(D-2)!}  \nn \\
&&+\f{1}{h}\e^{\l...\a_{D-2}\m\n} \Big (\f{a}{2}T^{\a_1}_{\ \ \m\n} +aT^{\ \a_1}_{\m\ \ \n}+cT^{\th}_{\ \m\th} g^{\a_1}_{\n}\Big )\cdot a\cdot\f{1}{(D-3)!} \nn \\
&&+\f{1}{h}\e^{\a_1\th\a_3...\a_{D-2}\m\n} \Big (\f{a}{2} T_{\th\m\n} +a T_{\m\th\n}+cT^{\r}_{\ \m\r} g_{\th\n}\Big )g^{\l\a_2}\cdot c\cdot\f{1}{(D-3)!} \bigg].
\eea
The coefficients $\f{1}{(D-2)!}$, $\f{1}{(D-3)!}$ are included here to remove the equivalent terms of the summation.

From the basic formula (\ref{23}), we see that for any 2-form in any dimensional Lorentzian manifold,
two operations of Hodge dual always result in minus the 2-form, i.e. $(\st \st T)^{\l}_{\ \m\n} = - T^{\l}_{\ \m\n}$.
Through a similar manipulation as in the previous sections, we arrive at:
\be
\label{66}
\begin{cases}
2a^2-ac = 1, \\
2a^2+ac = 0, \\
2a^2+(D-3)ac = 0.
\end{cases}
\ee
Now we have three equations for two unknown quantities.
One may observe that this equation set generally has $no$ solution unless in the four dimensional case, thus the original definition (\ref{h4.1}) could not be directly extended to a spacetime in general dimensions.

\subsection{A modified approach}            


To explore the possibility of a new generalized Hodge dual, let's first notice the fact that the action of teleparallel gravity described in language of differential forms involves only one Hodge dual operation of the field strength. Meanwhile, its Lagrangian is known to assume the following form
\be
\label{2234}
\mathcal{L}=h \Big(\f{1}{4}T_{\l\m\n}T^{\l\m\n}+\f{1}{2}T_{\l\m\n}T^{\m\l\n}-T^{\th}_{\ \m\th}T^{\r\m}_{\ \ \ \r} \Big).
\ee
The form of this action is actually independent of the spacetime dimension, see e.g. \cite{Ortin:2004ms}.
To ensure the equivalence to general relativity, the coefficients in (\ref{61}) are uniquely determined to be $a=\f{1}{2}, c=-1$
which are the same as the $D=3+1$ case.
In other words,
the first Hodge dual operation on torsion tensor in teleparallel gravity must be defined as
\be
\label{67}
(\st T)^{\l}_{\ \a_1...\a_{D-2}}=h\e_{\a_1...\a_{D-2}\m\n} \Big (\f{1}{4}T^{\l\m\n} +\f{1}{2}T^{\m\l\n}-T^{\th\m}_{\ \ \th} g^{\l\n}\Big ).
\ee
Note that the action (\ref{2234}) has been used as a starting point for the recent investigation of five dimensional teleparallel gravity \cite{Geng:2014yya}.

Let's now recall another constraint imposed by mathematical consistency that two successive operations of generalized Hodge dual should map torsion tensor back to itself, i.e. the key property (\ref{23}). In order to satisfy this condition and retain (\ref{67}) at the same time, the second operation of generalized Hodge dual must be defined appropriately. While this is trivially realized in $D=3+1$ where the second Hodge dual has the same parameters as the first Hodge dual, its realization in a spacetime with general dimensions is not so straightforward.

In order to find the second operation of generalized Hodge dual, it is convenient to introduce three new coefficients: $A, B$ and $C$.
We also denote the second Hodge dual operation with a new sign $*$ to distinguish it
from the first Hodge dual operation.
If we reserve the original coefficients $a,c$ for the moment, two operations of the Hodge dual lead to the following expression
\bea
\label{68}
(* \st T )^{\l}_{\ \a\b}&=&[* (\st T)]^{\l}_{\ \a\b}  \nn \\
&=&h\e_{\a\b\a_1...\a_{D-2}}\bigg [(\st T)^{\l\a_1...\a_{D-2}}\cdot a \cdot \f{1}{(D-2)!}\cdot A+(\st T)^{\a_1\l\a_2...\a_{D-2}}\cdot a\cdot\f{1}{(D-3)!}\cdot B \nn \\
&&+(\st T)_{\th}^{\ \a_1\th\a_3...\a_{D-2}}g^{\l\a_2}\cdot c\cdot\f{1}{(D-3)!}\cdot C \bigg ] \nn \\
&=&h\e_{\a\b\a_1...\a_{D-2}}\bigg [\f{1}{h}\e^{\a_1...\a_{D-2}\m\n} \Big (\f{1}{2}aT^{\l}_{\ \m\n} +aT^{\ \l}_{\m\ \n}+cT^{\th}_{\ \m\th} g^{\l}_{\n}\Big )\cdot a \cdot \f{1}{(D-2)!}\cdot A  \nn \\
&&+\f{1}{h}\e^{\l...\a_{D-2}\m\n} \Big (\f{1}{2}aT^{\a_1}_{\ \ \m\n} +aT^{\ \a_1}_{\m\ \ \n}+cT^{\th}_{\ \m\th} g^{\a_1}_{\n}\Big )\cdot a\cdot\f{1}{(D-3)!}\cdot B \nn \\
&&+\f{1}{h}\e^{\a_1\th\a_3...\a_{D-2}\m\n} \Big (\f{1}{2}a T_{\th\m\n} +aT_{\m\th\n}+cT^{\r}_{\ \m\r} g_{\th\n}\Big )g^{\l\a_2}\cdot c\cdot\f{1}{(D-3)!}\cdot C \bigg ].
\eea
After a similar calculation as before, we get
\be
\label{69}
\begin{cases}
a^2(A+B)-ac\cdot C = 1, \\
a^2(A+B)+ac\cdot C = 0, \\
-ac\cdot A+\big[(D-2)ac+2a^2\big]\cdot B = 0.
\end{cases}
\ee
It is reassuring that a solution could indeed be found once we insert the values of $a, c$.
Concretely the results are
\be
\label{70}
\begin{cases}
A = \f{2(D-3)}{D-2}, \\
B = \f{2}{D-2}, \\
C = 1.
\end{cases}
\ee

To sum up, in order to be compatible with the physical requirement that the teleparallel gravity is equivalent to general relativity,
and the mathematical requirement that two successive operations of generalized Hodge dual map the torsion tensor back to itself,
the generalized Hodge dual for torsion tensor in teleparallel gravity must be defined twofold.
The first operation of generalized Hodge dual is defined as
\be
\label{71}
(\st T)^{\l}_{\ \a_1...\a_{D-2}}=h\e_{\a_1...\a_{D-2}\m\n} \Big (\f{1}{4}T^{\l\m\n} +\f{1}{2}T^{\m\l\n}-T^{\th\m}_{\ \ \th} g^{\l\n}\Big ),
\ee
while the second operation of generalized Hodge dual must be defined in a spacetime-dimension dependent way as
\bea
\label{72}
(* \st T )^{\l}_{\ \a\b}&=&[* (\st T)]^{\l}_{\ \a\b} \nn  \\
&=&h\e_{\a\b\a_1...\a_{D-2}}\bigg [\f{1}{(D-2)^2(D-4)!}\cdot (\st T)^{\l\a_1...\a_{D-2}} \nn  \\
&&+\f{1}{(D-2)!}\cdot  (\st T)^{\a_1\l\a_2...\a_{D-2}}-\f{1}{(D-3)!}\cdot (\st T)_{\th}^{\ \a_1\th\a_3...\a_{D-2}}g^{\l\a_2} \bigg ].
\eea
Note that for a general Euclidean spacetime, each of the basic formulas  (\ref{23}) and  (\ref{999}) involves an extra sign, so we will essentially get the same results as the Lorentzian case.

This quite intriguing or even doubtful discovery may deserve further investigation.
However, building on our previous discussions, 
we may safely conclude that the approach of Lucas and Pereira is only applicable in four dimensional teleparallel gravity.

\section{Conclusion}      \la{con}

As a gauge theory for the translation group, teleparallel gravity deserves a Hodge dual for its gauge field strength, the torsion tensor, as that always happens in usual internal gauge theories. However, that the internal and external indices in this theory can be transformed into each other, which leads to appearance of new terms in the possible Hodge dual, makes introducing a generalized Hodge dual operator a nontrivial work.

In the context of four dimensional teleparallel gravity, such a definition for the torsion tensor has been proposed in \cite{Lucas:2008gs}. At first glance, it seems that a well defined generalized Hodge dual in general dimensions can be reasonably expected. However,
we find that a naive implementation of their approach  will inevitably lead to
the conclusion that no generalized Hodge dual exists in general dimensions, see (\ref{50}) and  (\ref{66}). Since the Lagrangian of teleparallel gravity takes a form which is independent of spacetime dimension, a key property of any consistent Hodge dual demands a new definition for the second operation of a generalized Hodge dual in general dimensions. After some exploration,
this possible generalization for the torsion tensor has finally been obtained, see  (\ref{71}) and  (\ref{72}).
Although this discovery definitely demands an explanation, we have no wisdom to add at this moment. Furthermore, these
results imply that the approach of Lucas and Pereira can only be implemented in dimension four,
but not in the general case.

One should notice that the form of the new generalized Hodge dual presented here is model dependent. If we take New General Relativity \cite{Hayashi:1979qx}, a generalized teleparallel model with three arbitrary parameters that $\mathcal{L}=h \Big(a_1T_{\l\m\n}T^{\l\m\n}+a_2T_{\l\m\n}T^{\m\l\n}+a_3T^{\th}_{\ \m\th}T^{\r\m}_{\ \ \ \r} \Big)$, instead of teleparallel gravity considered in this work, one would apparently 
arrive at new constraint equations,
and get a different definition of the generalized Hodge dual for torsion tensor.
Nevertheless, 
one could still use the modified approach to find a \textit{possibly} consistent generalized Hodge dual.

For further directions, the generalized Hodge dual for Riemann tensor in general dimensions may be investigated along the same line
(which deserves a separate treatment).
Furthermore,
in view of the discussion in \cite{deAndrade:2005xy},
the relevant self dual teleparallel gravity is worth of further research.
Thirdly, 
it may be possible to use the generalized Hodge dual to reproduce
the Lagrangian of conformally invariant teleparallel gravity \cite{{Maluf:2011kf},{Haghani:2012bt},{Bamba:2013jqa},{Formiga:2013iv}}.
Finally,
our work may have interesting connections with some other previous studies in \cite{{Obukhov:2002tm},{daRocha:2009sq},{Kofinas:2014owa},{Geng:2014yya}}.


\end{document}